\documentclass[prl, twocolumn, aps,amsmath,amssymb,amsfonts, superscriptaddress]{revtex4}
\usepackage{graphicx}% Include figure files
\usepackage{graphics}% Include figure files
\usepackage{dcolumn}% Align table columns on decimal point
\usepackage{multirow}
\usepackage{bm}% bold math
\usepackage{latexsym}
\usepackage{amssymb}
\usepackage{amsmath}
\usepackage{amsfonts}
\usepackage{layout}
\usepackage{verbatim}
\usepackage{epsfig}
\usepackage{amsbsy}
\usepackage[toc]{appendix}
\usepackage{xcolor}

\newcommand{\bea}{\begin{eqnarray*}}
	\newcommand{\eea}{\end{eqnarray*}}
\newcommand{\bne}{\begin{equation*}}
\newcommand{\ede}{\end{equation*}}

\newcommand{\bnen}{\begin{equation}}
\newcommand{\eden}{\end{equation}}
\newcommand{\bean}{\begin{eqnarray}}
\newcommand{\eean}{\end{eqnarray}}
\newcommand{\bsen}{\begin{subequations}}
	\newcommand{\esen}{\end{subequations}}

\newcommand{\bna}{\begin{array}}
	\newcommand{\eda}{\end{array}}
\newcommand{\bnm}{\begin{enumerate}}
	\newcommand{\edm}{\end{enumerate}}

\usepackage[utf8]{inputenc}

\begin{document}

\title{Non-linear anomalous Hall effect of two-dimensional spin-3/2 heavy holes}
\author{Sina Gholizadeh}
\affiliation{School of Physics, The University of New South Wales, Sydney 2052, Australia}
	\affiliation{ARC Centre of Excellence in Future Low-Energy Electronics Technologies, The University of New South Wales, Sydney 2052, Australia}
\author{Dimitrie Culcer}
\affiliation{School of Physics, The University of New South Wales, Sydney 2052, Australia}
	\affiliation{ARC Centre of Excellence in Future Low-Energy Electronics Technologies, The University of New South Wales, Sydney 2052, Australia}
\begin{abstract}
We identify a sizable non-linear anomalous Hall effect in the electrical response of spin-3/2 heavy holes in zincblende semiconductor nanostructures. The response is driven by a quadrupole interaction with the electric field enabled by $T_d$-symmetry. This interaction, until recently believed to be negligible, reflects inversion symmetry breaking and in two dimensions results in an electric-field dependent correction to the in-plane $g$-factor. The effect can be observed in state-of-the-art heterostructures, either via magnetic doping or by using a vector magnet, where even for small perpendicular magnetic fields it is comparable in magnitude to topological materials. 
\end{abstract}
%We identify a sizable non-linear anomalous Hall effect in the electrical response of spin-3/2 heavy holes in zincblende nanostructures. The response is driven by a quadrupole interaction with the electric field enabled by Td-symmetry, and until recently it was believed to be negligible, while in 2D results in an electric-field dependent correction to the in-plane g-factor. The effect can be observed in state-of-the-art heterostructures, either via magnetic doping or by a vector magnet, where even for small perpendicular magnetic fields it is comparable in magnitude to topological materials.
\date{\today}

\maketitle

% Referees: Menno, Giordano, Leonid Golub

\textit{Introduction} - Recent years have seen a surge in interest in non-linear electromagnetic responses motivated by outstanding advances in topological materials and semiconductor growth \cite{boyd2020nonlinear, morimoto2016topological, watanabe2021chiral, culcer2020transport, dobardvzic2015generalized}. Non-linear optical responses such as second-harmonic generation \cite{hipolito2017second, golub2014valley, golub2011valley, gao2021current}, shift currents \cite{sipe2000second, nakamura2017shift, rangel2017large}, the circular photogalvanic \cite{zhang2018photogalvanic} and resonant photovoltaic effects \cite{bhalla2020resonant} are being explored for technological applications including AC to DC conversion, photo-detection and energy harvesting. At the same time, non-linear electrical responses have revealed the existence of novel physical phenomena such as the anomalous Hall effect in time-reversal preserving systems \cite{PhysRevLett.115.216806}, which was recently observed in topological materials \cite{du2018band, ma2019observation, kang2019nonlinear}.

Second-order electrical responses require inversion symmetry breaking \cite{tzuang2014non, tokura2018nonreciprocal, gao2019nonreciprocal, shao2020non}. Aside from most topological materials, which have been at the heart of this effort, tetrahedral semiconductors likewise break inversion symmetry. This symmetry breaking is strong in zincblende crystals such as GaAs, and is associated with the spin-orbit interaction, which is particularly large in spin-3/2 hole systems. The effective spin-3/2 makes holes qualitatively different from spin-1/2 electrons \cite{luttinger1956quantum, chow1999semiconductor, winkler2003spin, cardona2005fundamentals, winkler2008spin, wang2010spin, biswas2014wave, Mawrie2014, Durnev_PRB2014, shanavas2016theoretical, akhgar2016strong, marcellina2017spin, Mawrie_JPCM2017, marx2020spin, wang2022ultrafast, bosco2021hole, froning2021strong, biswas2015zitterbewegung}, endowing them with unconventional properties such as a density-dependent in-plane $g$-factor \cite{miserev2017dimensional, marcellina2018electrical}, a strong anisotropy in both of the longitudinal conductivity and the Hall coefficient $R_H$ \cite{liu2018strong, marcellina2020signatures}, a non-monotonic Rashba spin-orbit coupling \cite{wang2021optimal}, a planar anomalous Hall effect \cite{cullen2021generating}, and superconductivity \cite{Hendrickx2018}. Until recently tetrahedral $T_d$ symmetry terms were believed to be negligible in hole systems \cite{winkler2003spin}, yet a more careful evaluation has demonstrated their size to be significant \cite{philippopoulos2020pseudospin}, so that sizable second-order electrical responses should be possible in hole systems.

% I plotted for two different values of M only to compare them. 
\begin{figure}[tbp]
    \includegraphics[width=\columnwidth]{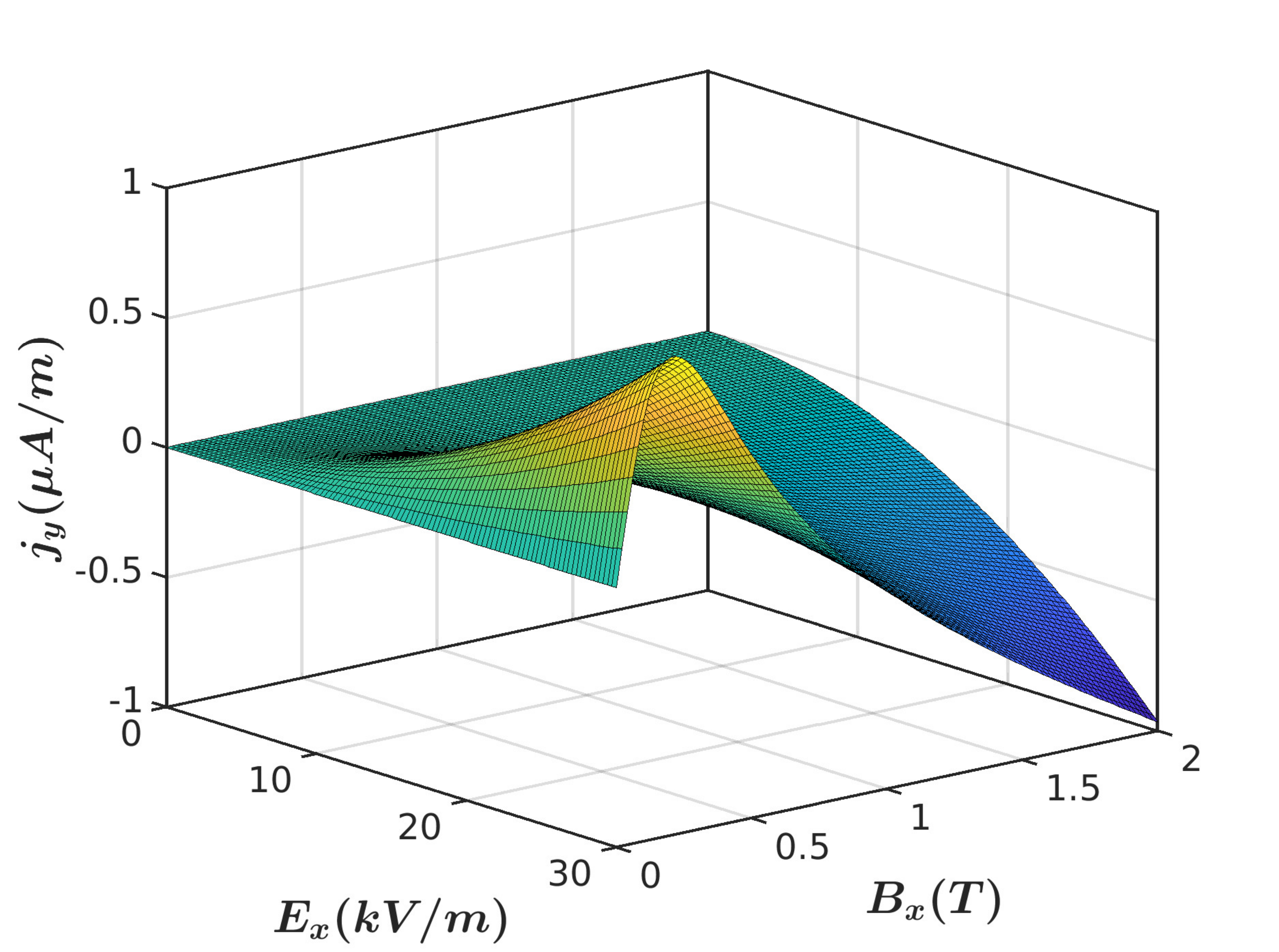}
    \caption{Non-linear anomalous Hall current density $\parallel \hat{\bm{y}}$ as a function of the in-plane magnetic field and of the driving electric field for a $30 \, nm$ wide GaAs quantum well. The perpendicular Zeeman energy is $M = 0.1$ meV.}
    \label{fig:3D}
\end{figure}

\begin{figure}[tbp]
    \includegraphics[width=\columnwidth]{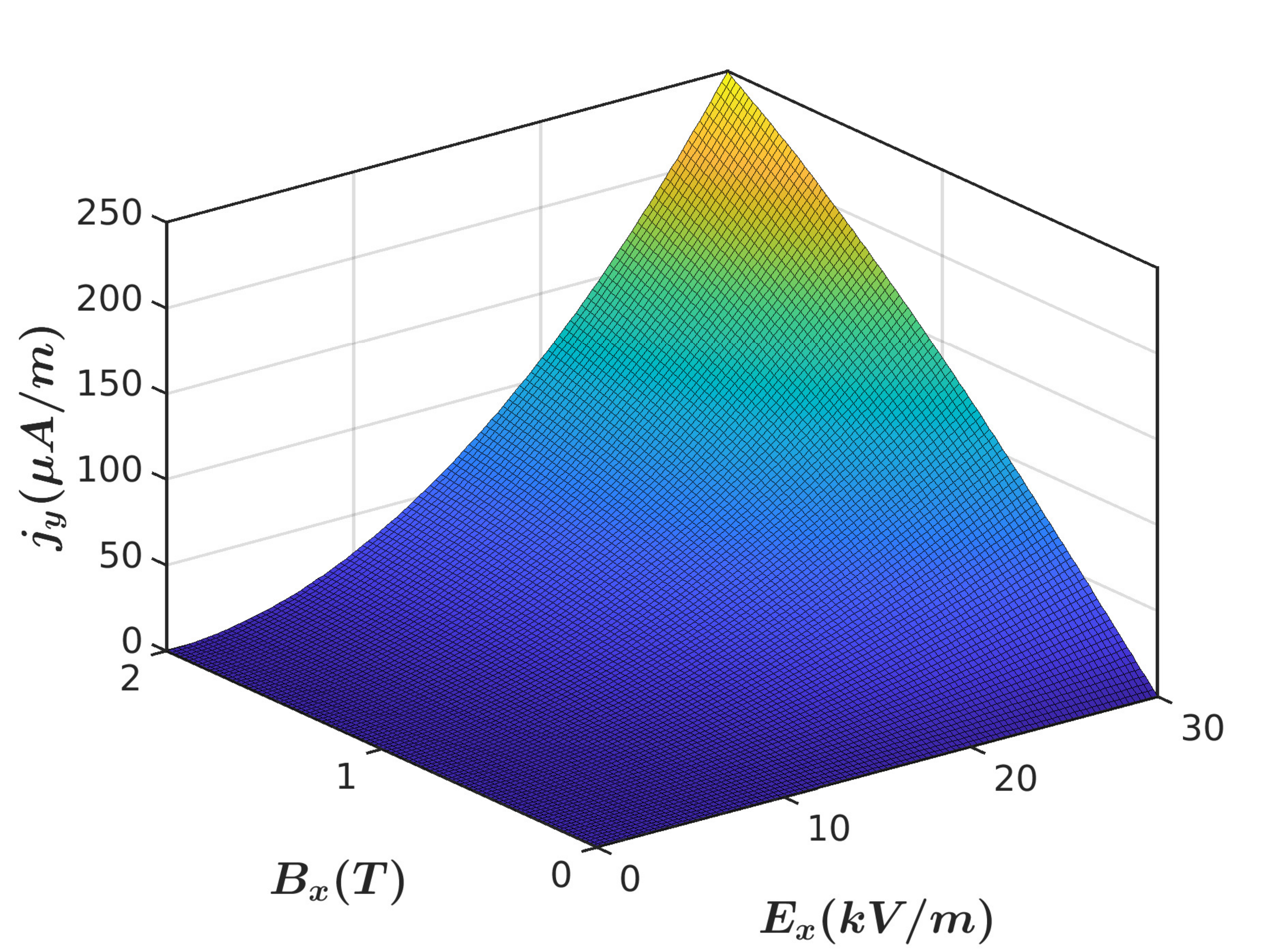}
    \caption{Non-linear anomalous Hall current density $\parallel \hat{\bm{y}}$ as a function of the in-plane magnetic field and of the driving electric field for a $30 \, nm$ wide GaAs quantum well. The perpendicular Zeeman energy is $M = 2$ meV.}
    \label{fig:3D_2nd}
\end{figure}

In this work, we show that tetrahedral symmetry terms lead to a non-linear anomalous Hall effect in a symmetric hole quantum well. The effect occurs in the presence of both in-plane and out-of-plane Zeeman fields, the latter of which can be produced either by a small magnetic field or by magnetic impurities in a ferromagnetic semiconductor. The role of the $T_d$ terms can be understood as an electric-field induced shear term in the in-plane $g$-factor. Our main result is summarized in Figs.~\ref{fig:3D} and \ref{fig:3D_2nd}. We find a sizable non-linear anomalous Hall current density along the $y$-axis, accompanied by a much smaller non-linear longitudinal current density, not shown. The effect can be easily measured in readily available hole nanostructures, which provide a straightforward set-up for probing the existence of tetrahedral-symmetry terms. Based on realistic parameters we find that the non-linear anomalous Hall effect in spin-3/2 holes can be comparable in magnitude to the values reported recently in topological materials \cite{kang2019nonlinear}. Aside from the novelty of identifying a non-linear electrical response purely due to holes, as opposed to well-known optical transitions linking the valence and conduction bands, a sizable tetrahedral contribution beyond the Luttinger model will have important repercussions for hole-based quantum computing, \cite{Nichele-Ensslin-2014-PRB, salfi2016charge, Brauns-Zwanenburg-2016-PRB, Matthias-Zwanenburg-2016-APL, Daisy-2016-NL, Katsaros-2016-NL, Salfi-2016-Nanotechnology, Nichele-Kouwenhoven-2017-PRL, Srinivasan-2017-PRL, Conesa-Boj-2017-NL, Li_APL2017, Jo-2017-PRB, Liles2018, Vukusic_NL2018, Li_NL2018, Silvano_PRL2018, Hendrickx2018, Hendrickx2020} where it may enable additional possibilities for the electrical manipulation of holes.

% HOLE QUANTUM COMPUTING: 
% Cite 2-3 recent papers by Daniel Loss (2019-2021). 
% arXiv:2003.07079
% https://www.nature.com/articles/s41467-021-27880-7

% Fig. 1 E_x from 0 up to 3x10^4 V/m.
% Format captions and update numbers in captions. 
% Get rid of all j_x plots.
% Update Fig. 5. Also I // 100.

% In the discussion you need to discuss the shape of ALL the plots: 2, 3, 4. 
% Write up the estimate for E_y and V_y and explain how you got it. Must be mentioned in the main text under Exptl observation. 

\textit{Hamiltonian}. We consider a symmetric hole quantum well grown in a zinc blende heterostructure along the high-symmetry crystallographic direction (001). The confinement is along the $z$-axis. For concreteness we consider GaAs with Luttinger parameters for $\gamma_1 = 6.85$, $\gamma_2 = 2.10$ and $\gamma_3 = 2.90$. The total Hamiltonian $H = H_0 + U + H_E$ includes the band Hamiltonian, the disorder potential and the applied electric field. The band Hamiltonian is the combination of kinetic part and $H_Z$, $H_0  = \hbar^2 k^2/2 m+ H_Z$ where $H_Z$ is given by,
\begin{align}
& H_Z = \Delta _1 B_+ k_+^2 \sigma _- +\Delta _2 B_- k_+^4 \sigma _- + \textbf{\textit{h.c.}} +M \sigma _3
\end{align}
where $\Delta_1$ and $\Delta_2$ are the g-factors, $B_{\pm} = B_x \pm i B_y$ represent the in-plane magnetic field of magnitude $B_\parallel$, $\textbf{\textit{h.c.}}$ is Hermitian conjugate, $k_{\pm} = k_x \pm i k_y$, while $\sigma_{\pm} = (1/2) \, (\sigma_x \pm i \sigma_y)$. Without loss of generality we set $B_\parallel = B_x$. We have denoted the Zeeman field in the $\hat{\bm z}$-direction by $M$, and define it so that $M$ has units of energy. The eigenvalues of $H_0$ are $\epsilon _{\bm{k}}^{(\pm)} = \epsilon _{k}^{(0)} \pm \Omega_{\bm{k}}$ where $\epsilon_k^{(0)} = \hbar^2 k^2/2 m$ is kinetic part and energy dispersion is split by $ \Omega_{\bm{k}} = \sqrt{M^2+B_\parallel^2 k^4 G_{\bm{k}}} $, and $ G_{\bm{k}}= (\Delta_1)^2 + k^4 (\Delta_2)^2 +2 k^2 \Delta_1\Delta_2 \cos (2 \theta)$.

The two-by-two Hamiltonian given by $H_0$ is the projection of Luttinger Hamiltonian under Schrieffer-Wolff transformation to HH subspace, and the momentum-dependent $g$-factors reflect the strong spin-orbit interaction in Luttinger Hamiltonian \cite{chow1999semiconductor, winkler2003spin}. The interaction is known to be proportional to $B_\parallel k^2$, a term quadratic in in-plane momentum \cite{winkler2003spin} while recent work has identified a new term $B_\parallel k^4$ \cite{miserev2017dimensional} quartic in in-plane momentum, which has been confirmed experimentally \cite{miserev2017mechanisms}. Such momentum-dependent Zeeman terms are specific to heavy holes, which represent the $\pm 3/2$ projection of the hole spin-$3/2$ onto the quantization axis. Note that, although the electric field is treated perturbatively, the Zeeman terms are treated exactly. The prefactors $\Delta_1$ and $\Delta_2$ are functions of $k$ and decrease strongly at larger wave vectors. Such terms are not present in spin-$1/2$ electron systems. In our evaluations this $k$-dependence is taken into account: we evaluate the coefficients for the specific carrier density we have chosen. The Zeeman interaction for heavy holes also includes terms $\propto B_\parallel^3$, yet these are three orders of magnitude smaller than the $B_{\parallel}-$linear terms above, only becoming important for $B_{\parallel} \ge 30$ T.

For a symmetric quantum well there is no Rashba spin-orbit interaction \cite{winkler2003spin, moriya2014cubic}. Likewise, we have not included the Dresselhaus interaction in the Hamiltonian $H_0$ because it does not contribute to the non-linear signal: we have verified this explicitly. The interaction with a uniform electric field $\bm{E}$ makes two contributions to $H_E$, namely $H_E = e \bm{E}. \hat{\bm{r}} + H_\lambda$. The first is the customary electrostatic potential, while the second, denoted by $H_{\lambda}$, arises from $T_d$ symmetry, and has the form
\begin{align}
    H_\lambda =  i e \lambda  M (E_+ k_-^2 \sigma_+ - E_- k_+^2 \sigma_-)
    \label{H_lambda_1st}
\end{align}
in which $E_{\pm}$ is the in-plane electric field, $M$ is the out-of-plane Zeeman field which we take to have units of energy, and $\lambda$ is a material-specific parameter whose magnitude will be determined below. We transform $H_{\lambda}$ to the eigenstate basis of $H_0$. Without loss of generality we assume that the electric field points in the $\hat{\bm{x}}$ direction, yielding $\tilde{H}_{\lambda} = e E_x k^2 M \lambda \tilde{\sigma}_y$, where the tilde indicates matrices in the eigenstate basis, hence $\tilde{\sigma}_y$ is a pseudospin rather than a spin matrix. By comparing $H_\lambda$ and $H_Z$ we note that $H_\lambda$ can be understood as an electric-field correction to the Zeeman Hamiltonian: $T_d$-symmetry mixes the out-of-plane Zeeman field with the in-plane electric field, yielding an additional in-plane Zeeman interaction. This is essentially an electrically tunable shear term in the $g$-tensor. An analogous correction is present for the out-of-plane Zeeman interaction due to an out-of-plane electric field and an in-plane Zeeman field, as discussed in the Supplement.

\begin{figure}[tbp]
    \includegraphics[width=0.7\linewidth]{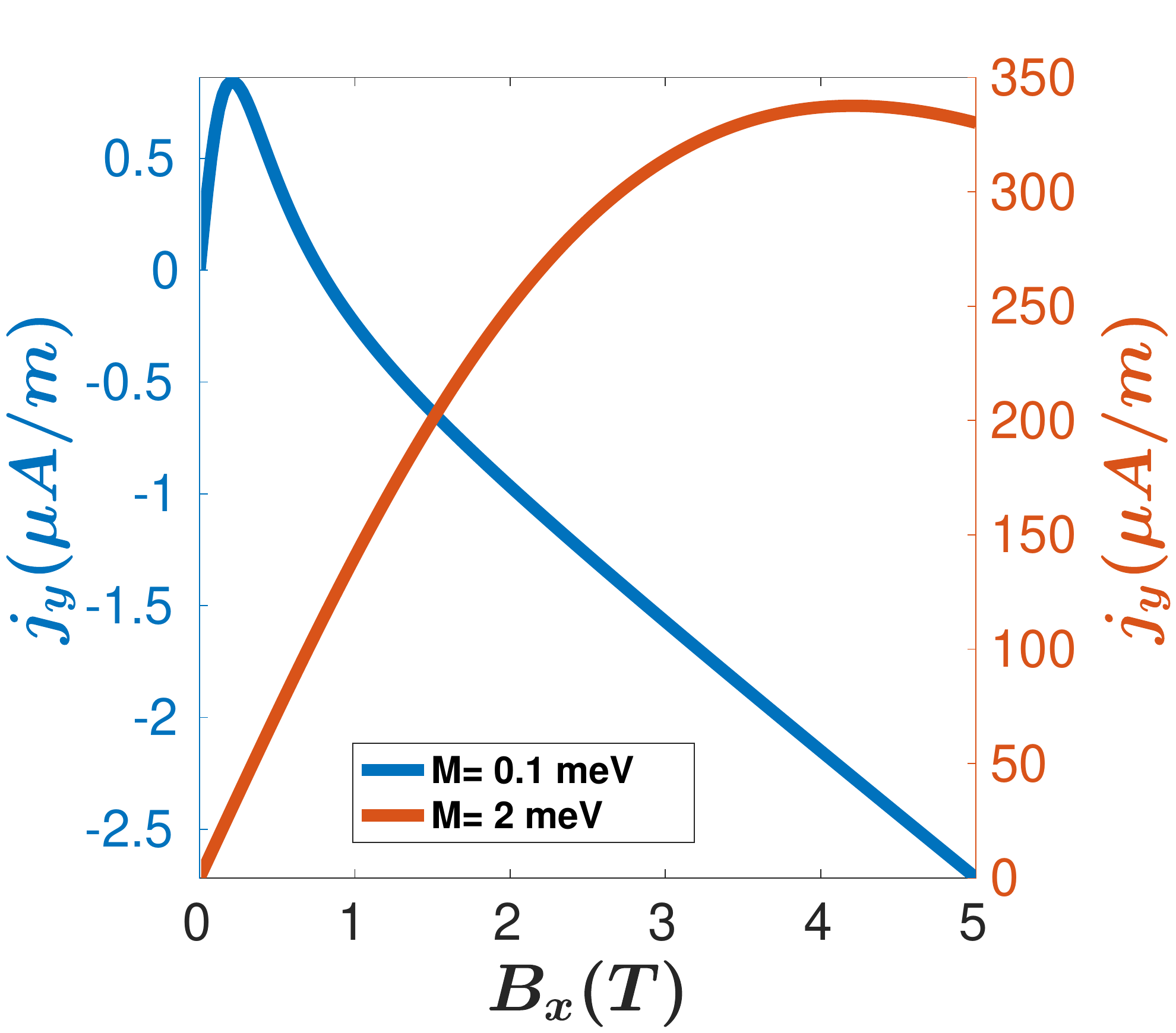}
    \centering
    \caption{The non-linear anomalous Hall effect is a non-monotonic function of the in-plane magnetic field, shown here for two different values of the out-of-plane Zeeman field: $M=2$ meV and $M=0.1$ meV. The electric field is $E_x = 30 \, kV/m$.} 
    \label{fig:plot_B_B}
\end{figure}

% This is of course not the quantum Liouville equation. 

\textit{Kinetic equation}. The quantum Liouville equation for first order in electric field is given by,
\begin{equation}
    \frac{\partial \rho_E}{\partial t} + \frac{i}{\hbar}[H_0,\langle \rho_E \rangle] + J(\langle \rho_E \rangle) = - \frac{i}{\hbar}[H_E,\langle \rho_0 \rangle]
    \label{differential_1st}
\end{equation}
where $\rho_0$ and $\rho_E$ are the equilibrium and first-order response density matrices. Scattering term in the Born approximation is written as
$J(\langle \rho \rangle) = \frac{1}{\hbar^2}\int_{0}^{\infty} dt' \langle [U,[e^{-i H_0 t'/\hbar} U e^{i H_0 t'/\hbar},\langle \rho(t) \rangle ]] \rangle$. We assume that the disorder potential is given by a short-range disorder $U(\bm{r})=U_0 \sum_i \delta (\bm{r}-\bm{r}_i)$ and the correlation function yields us $\langle U(\bm{r}) U(\bm{r}') \rangle = n_i U_0^2 \, \delta(\bm{r}- \bm{r}')$ with $n_i$ the impurity density. The driving term has two contributions,
\begin{align}
	D_{E\bm{k}}= \frac{e E_x}{\hbar} \bigg\{\frac{\partial \rho_0}{\partial k_x} - i \left[\mathcal{R}_{k_x},\rho_0 \right] \bigg\}-\frac{i}{\hbar}[\tilde{H}_{\lambda },\rho _0]
    \label{driving_term}
\end{align}
in which $\mathcal{R}_{k_x} = \langle u_{\bm{k}} | \left(i \partial u_{\bm{k}}/ \partial k_x\right)  \rangle$ is the Berry connection. In the crystal momentum representation, i.e. $|m,\bm{k} \rangle=e^{i\bm{k}.\bm{r}} |u_{m \bm{k}} \rangle$, the equilibrium density matrix is simply Fermi-Dirac distribution function given at energy $\epsilon_{\bm{k}}^{(m)}$, $\langle m, \bm{k} | \langle \rho_0 \rangle | m', \bm{k} \rangle = f_0 (\epsilon_{\bm{k}}^{(m)}) \delta^{m m'}$. The eigenstates are evaluated in the absence of the electric field. The terms in the curly brackets are the consequences of $e \bm{E}. \hat{\bm{r}}$. The first term in curly bracket is the Fermi surface response. The second term in curly bracket and the term outside the curly bracket contains Fermi sea responses. 

To determine the density matrix response $\rho_E$ we solve equation \ref{differential_1st} up until a linear order of the electric field. Using the same differential equation we try to find the second-order density matrix responses but this time we solve the equation followed by these substitutions, $\rho_0 \rightarrow \rho_E$ and $\rho_E \rightarrow \rho_{E2}$. Having access to driving term we can derive the diagonal and off-diagonal parts of the density matrix \cite{culcer2017interband}. The off-diagonal part of the density matrix $\rho_E$ can be calculated using the off-diagonal terms in the driving term (Eq. \ref{driving_term}, refer to appendix). Accordingly, the band-diagonal part of the density matrix is given by
\begin{align}
    J_{d}[(n_E)^{(-1)}]^{m m}= \frac{e \bm{E}}{\hbar} \cdot \frac{\partial f_0(\epsilon_{\bm{k}}^{(m)})}{\partial \bm{k}}
    \label{J_d(m,m)}
\end{align}
where $ \partial f_0 (\epsilon_{\bm{k}}^{(m)})/\partial \epsilon_{\bm{k}}^{(m)} \approx -\delta  [\epsilon_{\bm{k}}^{(m)} - \epsilon _F]$ at low temperatures, with $\epsilon _F$ the Fermi energy.

%\begin{figure}[tbp]
%    \includegraphics[width=0.7\linewidth]{fig/plot_E.eps}
%    \centering
%    \caption{Current density along y-axis in second-order response behaves parabolically in terms of the electric field. Magnetic field and perpendicular Zeeman energy are considered as, $B=1 \, T$, and $M = 0.1 \, meV$, respectively. }
%    \label{fig:plot_E_E}
%\end{figure}

\begin{figure}[tbp]
    \includegraphics[width=0.7\linewidth]{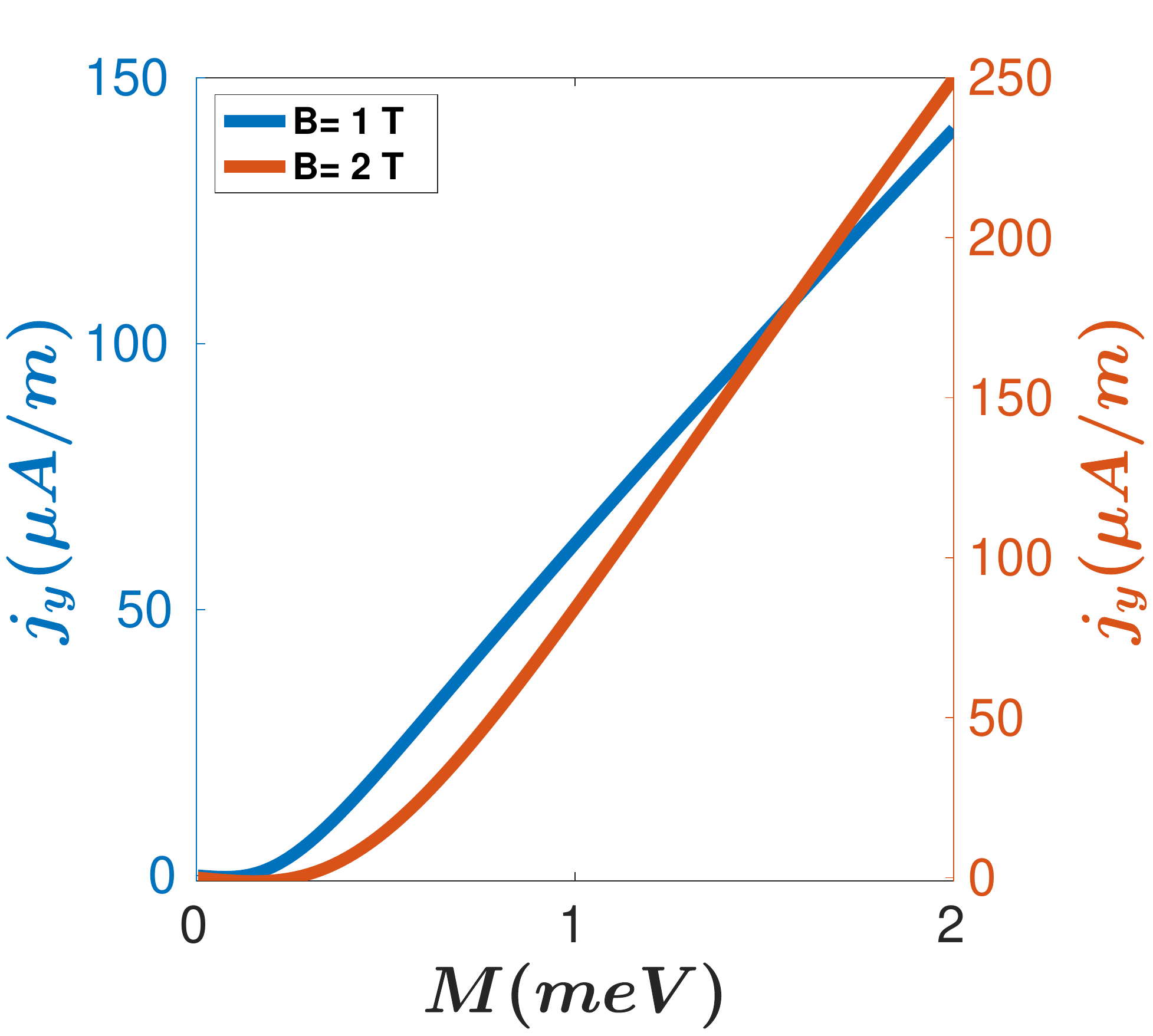}
    \centering
    \caption{Non-linear anomalous Hall current under the considered range of values for the parameters in current densities behave approximately linearly in terms of magnetization ($M$). Electric field is considered as $E_x=30 \, kV/m$. }
    \label{fig:plot_M_M}
\end{figure}

\textit{Second-order electrical response}. We analytically derive the conductivity with $\Delta_2$ set to zero. For pedagogical purposes we perform an analytical calculation first, using a simplified model. $ G_{\bm{k}}$ becomes simply $\Delta_1^2$ if we neglect $\Delta_2$. The term $\Omega_{\bm{k}}$ is function of both $k$ and $\theta$ if we take into account both g-factors. But if we only consider one of the $g$-factors the dispersion is isotropic. Hence, the overall calculation becomes much simpler if we only consider one of the $g$-factors. To second order we find $j_x = \chi_{x x x} E_x^2$ and $j_y = \chi_{y x x} E_x^2$, where 
\begin{align}
    & \chi_{x x x}  = \frac{2 e^3  B_x^3 \Delta_1^3 k_F^6 m M^2 \lambda}{\pi  \hbar^3 \Omega_F^4} \nonumber
    \\
    & \chi_{y x x}  = \frac{2 e^3  \lambda m M \tau  B_x \Delta_1 k_F^2 \left( \Omega_F^2 + M^2 \right)}{\pi \hbar^4 \Omega_F^2},
    \label{nonlinear-current-density}
\end{align}
where $\Omega_F = \sqrt{M^2 + B_x^2 \Delta_1^2 k_F^4}$, $k_F$ is the Fermi wave vector, and the momentum relaxation time $\tau = \hbar^3/(n_i U^2 m)$. Numerically we extend our results for the general case while $\Delta_2$ effect is taken into account. The transverse non-linear current density, shown in Figs. \ref{fig:3D} - \ref{fig:plot_M_M}, is larger than its longitudinal counterpart (not shown) by several orders of magnitude. In both cases, in the presence of $\Delta_2$, the behavior of the current density is non-monotonic in terms of the magnetic field but in general $|j_y|$ increases linearly in terms of the magnetic field while the magnetic field is larger compared to magnetization. %Current density along y-axis increases almost linearly as a function of in-plane magnetic field while it's decaying along x-axis as magnetic field increases (Fig. \ref{fig:3D}). 
As we expect, the current density in second-order behaves as an increasing parabolic function in terms of electric field (Fig. \ref{fig:3D} - \ref{fig:3D_2nd}). In Fig. \ref{fig:plot_M_M} we can see that in the range of values set for magnetic and electric field, the current density behaves almost linearly as a function of magnetization.

To obtain $\lambda$, we start from the Luttinger Hamiltonian combined with the electric-dipole and Zeeman Hamiltonians, and we apply the Schrieffer-Wolff transformation to project the system to the HH subspace. This yields an effective $2 \times 2$ Hamiltonian of the same form as $H_0$. In the axial approximation $\lambda = (3 a_B \xi)/(8 \bar{\gamma} \langle k_z^2 \rangle^2 \mu)$,
in which $a_B$ is the Bohr radius, and $\xi$ is a parameter controlling the intensity of electric-dipole terms. Due to the confinement along the $z$-axis we know that $k_z$ is comparatively larger than in-plane wave vectors \cite{winkler2008spin}.  For the following values, $\xi = 0.2, \mu = \hbar^2/m_0 \; (m_0 \, \text{is the bare electron mass}), \bar{\gamma} = 2.5, k_z = (\pi/30 \; nm) $, we obtain $\lambda = 1.08 \times 10^{-6} {\rm m s^2kg^{-1}}$. 

\textit{Discussion}. % Discuss each figure. Shape and applicability of Fig. 2. 
The conductivities that we calculated in the second-order response are directly proportional to $\lambda$. We found that $\lambda$ is proportional to the size of quantum well. Therefore, this effect will be pronounced in larger wells. Furthermore, it is easy to see that the second-order response thoroughly is dependent on $\xi$, meaning that at $\xi=0$ (for crystals with a center of inversion symmetry) the second-order response vanishes. Marcellina et al. \cite{marcellina2020signatures} shows that in spin-3/2 system with two different $g$-factors (each have different winding numbers) there appears a sizable anisotropy in conductivities and Hall coefficient. We have shown that in the similar system a sizable nonlinear response can be probed using $H_\lambda$ (Eq. \ref{H_lambda_1st}). Note that, although the non-linear anomalous Hall effect is stronger in higher mobility systems, the ratio $j^{(2)}/\sigma_{xx}$ is independent of the mobility, playing the role of a nonlinear Hall coefficient.

In the absence of $\Delta_2$ the absolute value of the current density increases monotonically as a function of the out-of-plane Zeeman energy and in-plane magnetic field. When either quantity, i.e. $M$ or $B_x$, reaches a sizable value $|j_y|$ becomes essentially linear in that quantity. On the other hand, when $\Delta_2$, which has a different sign from $\Delta_1$, is accounted for in the Hamiltonian, $|j_y|$ is non-monotonic as a function of $B_x$ in the range of $M(meV)\in [0, 2]$ but it still behaves monotonically as a function of $M$ in the range of $B_x(T) \in [0, 2]$. Nevertheless, the current density behaves linearly at comparatively large (small) values of $M$ and small (large) values of $B_x$. One can also see from figure \ref{fig:plot_B_B} that the first derivative of $j_y$ changes sign depending on the size of the magnetization.

\begin{figure}[tbp]
    \includegraphics[width=\linewidth]{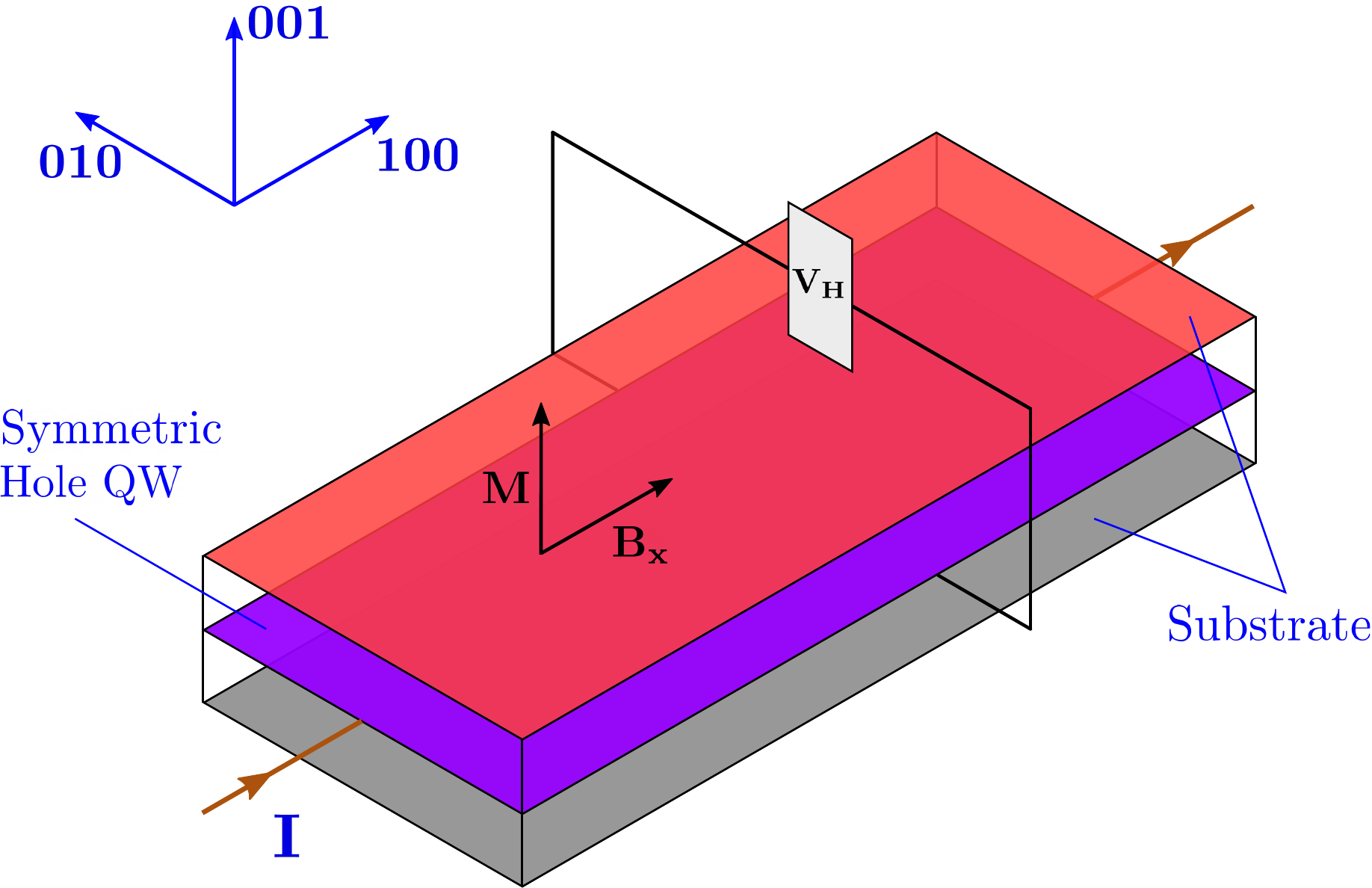}
    \centering
    \caption{Schematic of one potential experimental setup involving a symmetric hole quantum well. The effect can be detected by measuring the Hall voltage. }
    \label{fig:sample}
\end{figure}

\textit{Experimental observation}. There is considerable flexibility in regard to experimental observation, with one possibility illustrated in Fig.~\ref{fig:sample}. To detect a signal of second order in the applied electric field one requires a low-frequency alternating field, which can be accomplished using an oscillator with a frequency of the order of $\omega/(2\pi) \sim 100$ Hz. The Hall voltage at $2\omega$ can then be read out. The relaxation time $\tau$ in general is much smaller than the alternating electric field time period, i.e. $\omega \tau \ll 1$, hence our DC calculation is appropriate to describe this physics. The key feature is the way the perpendicular Zeeman field is generated: this can be accomplished either via a magnetic doping, as in  ferromagnetic semiconductors, or by applying a perpendicular magnetic field. We discuss each in turn. To facilitate comparison with nonlinear effects in topological materials we will use the exact sample size used in Ref.~\onlinecite{kang2019nonlinear}

The easiest way is using an unmagnetized GaAs sample together with a vector magnet generating an arbitrary magnetic field. The out-of-plane magnetic field can be set so that the Zeeman energy is $M\approx 0.1$ meV so that one can neglect the orbital terms giving rise to the ordinary linear Hall effect, even though these do not contribute to the non-linear anomalous Hall voltage. To determine the Hall voltage we start with the relationship between the resistivity and conductivity tensors
\begin{equation}
\left(\begin{array}{cc}
 \rho _{xx} & \rho _{xy} \\
 \rho _{yx} & \rho _{yy} \\
\end{array}
\right) = \frac{1}{\sigma _{xx}^2+\sigma _{yx}^2} \left(
        \begin{array}{cc}
         \sigma _{xx} & -\sigma _{xy} \\
         \sigma _{xy} & \sigma _{xx} \\
        \end{array}
        \right).
\end{equation}
The induced non-linear Hall electric field takes the form
\begin{align}
    E_y^{(2)} \approx \frac{ \chi _{yxx}}{\sigma _{xx}} E_x^2,
\end{align}
as outlined in the Supplement. Assuming $M=0.1$ meV, using figure \ref{fig:3D} we take the current density $0.96 \, \mu A/m$ corresponding to the magnetic field, electric field, and magnetization provided at $B_x = 2 \, T$, $E_x =30 \, kV/m$, and $M=0.1 \, meV$. To find the conductivity we consider that a relaxation time $\tau = 1$ ps. The hole carriers density is taken to be $1.86 \times 10^{15} \, m^{-2}$. The heavy hole in-plane mass is $m^* = m_0 / (\gamma_1+ \bar{\gamma})$, yielding
\begin{align}
    \sigma _{xx} = \frac{n e^2 \tau}{m^*} \approx 12.7 \, \frac{e^2}{h}, 
\end{align}
hence the induced non-linear Hall electric field $E_y^{(2)} \approx 2 \times 10^{-3} $ V/m and a corresponding non-linear Hall voltage $V_y^{(2)} = E_y^{(2)} d = 18$ nV. 

A class of state-of-art samples that can be used to study nonlinear effects are magnetic semiconductors, such as GaMnAs \cite{ohno1999electrical, ohno2000electric, fukumura2001magnetic, tang2004negative, gould2004tunneling, ruster2003very, giddings2005large, krainov2021spin}. The magnetization easy axis in GaMnAs can be either in the plane or out of the plane \cite{lee2009ferromagnetic}, and the orientation can be tailored by means of strain. We focus on the latter case in this example and assume a larger value of $M \approx 2$ meV. From figure \ref{fig:3D_2nd} we take the current density to be 0.25 mA/m corresponding to a magnetic field and electric field of $B_x = 2 \, T$ and E = 30 kV/m respectively. Using the same carrier density and relaxation time as above we find the same value for $\sigma_{xx}$, whereupon the induced non-linear Hall electric field $E_y^{(2)} \approx 0.5$ V/m, and, in a sample of width $9.2 \, \mu m$ as in \cite{kang2019nonlinear}, we find a Hall voltage $V_y^{(2)} \approx 5 \, \mu$V, which is comparable in size to Ref.~\onlinecite{kang2019nonlinear}.

\textit{In summary}, we have demonstrated that in a symmetric quasi-2D hole system a sizable non-linear anomalous Hall effect is present, driven by tetrahedral symmetry terms that go beyond the Luttinger model. The effect is measurable either in a conventional GaAs sample using a vector magnet, or in a sample of ferromagnetic GaAs in a magnetic field, where the magnetic field orientation is chosen depending on the direction of the magnetization. 

\acknowledgments

This work is supported by the Australian Research Council Centre of Excellence in Future Low-Energy Electronics Technologies (project number CE170100039). 

%\bibliography{main.bib}

\appendix
\begin{widetext}

\section{Appendix A: Basis states and scattering term}
% the \\ insures the section title is centered below the phrase: AppendixA

The eigenvector of $H_0  = h^2 k^2/2 m+ H_Z$ are calculated as,

\begin{align}
    u_{\bm{k}}^{(\pm)} = 
    \sqrt{\frac{\Omega_{\bm{k}} \mp M}{2 \Omega_{\bm{k}}}} \left(
        \begin{array}{c}
        \frac{e^{-2 i \theta } \left(M \pm \Omega_{\bm{k}} \right) }{B k^2 \left(\Delta_1+ e^{2 i \theta } k^2\Delta_2 \right) }  \\
        1 \\
        \end{array}
        \right)
    \label{eigenvector}
\end{align}
where $\Omega_{\bm{k}} = \sqrt{M^2+B_\parallel^2 k^4 G_{\bm{k}}} $, and $ G_{\bm{k}}= (\Delta_1)^2 + k^4 (\Delta_2)^2 +2 k^2 \Delta_1\Delta_2 \cos (2 \theta)$, while $\theta = \tan^{-1}\left( k_y / k_x \right)$. 

Since $H_{\lambda}$ is linear in the electric field, in addition to $e \bm{E}. \hat{\bm{r}}$, we need to add this term to driving term (Eq. \ref{driving_term}) too. To do so, We have to transform $H_{\lambda}$ using S matrix which is formed based on $H_0$ eigenvectors.
\begin{align}
	\tilde{H}_{\lambda} = S^{\dagger} H_{\lambda} S
\end{align}
where
\begin{align}
	S = \begin{pmatrix}
	u_{\bm{k}}^{(1)}(1,1) &  u_{\bm{k}}^{(2)}(1,1) \\
	u_{\bm{k}}^{(1)}(2,1) &  u_{\bm{k}}^{(2)}(2,1) 
	\end{pmatrix}. 
\end{align}
here $u_{\bm{k}}^{(1)}$, and $u_{\bm{k}}^{(2)}$ correspond to $u_{\bm{k}}^{(-)}$, and $u_{\bm{k}}^{(+)}$ (Eq. \ref{eigenvector}), respectively. Another parameter that we have used in driving term is the x-component of Berry connection (Eq. \ref{driving_term}). While $\Delta_2=0$, we find the Berry connection's x-component using the eigenvectors that we have calculated recently,
\begin{align}
    \mathcal{R}_{k_x} = \bigg \langle u_{\bm{k}} \bigg\rvert i \frac{\partial u_{\bm{k}}}{\partial k_x} \bigg \rangle = -\frac{\sin (\theta )}{k} \sigma_0 + \frac{M \sin (\theta )}{k \Omega_{k} } \sigma_3 +
    \frac{B \Delta_1 k \sin (\theta )}{\Omega_k } \sigma_1 
    - \frac{B \Delta_1 M k \cos (\theta ) }{ \Omega_k ^2} \sigma_2.
\end{align}
here, $\Omega_k$ is isotropic since $\Delta_2$ is considered to be zero.  
\\~\\
Off-diagonal part of the density matrix in linear and later in the second-order response can be derived using equation \ref{differential_1st}. Taking only the off-diagonal terms from both sides of equation \ref{differential_1st} leads us to
\begin{align}
    \frac{\partial S_{E\bm{k}}}{\partial t} + \frac{i}{\hbar} \left[H_0, S_{E\bm{k}}\right] = D_{E\bm{k}} + D'_{E\bm{k}}
    \label{differential_2nd}
\end{align}
Comparing both sides of equation \ref{differential_2nd} and using equation \ref{driving_term} we can write,
\begin{align}
    S_{E \bm{k}}^{m m'} = -i \hbar \mathcal{P} \left(\frac{D_{E \bm{k}}^{m m'}+ D_{E \bm{k}}^{\prime m m'}}{\epsilon_{\bm{k}}^{m}-\epsilon_{\bm{k}}^{m'}}\right)
    \label{S_E&D_E} 
\end{align}
while $D'_{E\bm{k}} = -J_{od}\left[n_{E\bm{k}}\right]$ (for further details please refer to \cite{culcer2017interband}). Off-diagonal part of density matrix takes a linear form in terms of electric field in linear response. In the case that $\Delta_2$ is set to zero we get for $S_{E\bm{k}}$,  
\begin{align}
    & S_{E\bm{k}} = \frac{e E_x k  B \Delta_1 \sin (\theta )}{2 \Omega_k^2 } \left(-f_1 + f_2\right) (\sigma_1) - \frac{e E_x k M (-B \cos (\theta ) \Delta_1 + k \lambda \Omega_k^2)}{2 \Omega_k ^3} \left(-f_1+f_2\right) (\sigma_2) 
\end{align}  
here $f_1$ and $f_2$ are Fermi-Dirac distribution function given at energies $\epsilon_{\bm{k}}^{(1)}$, and $\epsilon_{\bm{k}}^{(2)}$, respectively. $\Omega_k$ has no angular dependency, i.e., $\Omega_k = \sqrt{M^2 + B^2 \Delta_1^2 k^4}$. In our calculations, we see that $D_{E \bm{k}}^{\prime m m'}$ contribution to $S_{E\bm{k}}$ is zero. To get the diagonal part of the density matrix ($n_{E\bm{k}} = n_{E\bm{k}}^{(0)} \sigma_0 + n_{E\bm{k}}^{(\Omega)} \sigma_3$), we need to solve equation \ref{J_d(m,m)} along with the following equation \ref{J_d_diagonal_density}
\begin{align}
    [J_d(n)]_{\bm{k}}^{mm} = \frac{2 \pi n_i}{\hbar} \sum_{m' \bm{k}'} U_{\bm{k} \bm{k}'}^{m m'} U_{\bm{k}' \bm{k} }^{m' m} (n_{E \bm{k}}^{mm} - n_{E \bm{k}'}^{m' m'})  \delta(\epsilon_{\bm{k}}^m-\epsilon_{\bm{k}'}^{m'})
    \label{J_d_diagonal_density}
\end{align}
 which leads to  
\begin{align}
    [J_0(n_{E\bm{k}}^{(0)})]^{(1,1)} = [D_0]^{(1,1)} 
    \\
    [J_0(n_{E\bm{k}}^{(\Omega)})]^{(1,1)} = [D_\Omega]^{(1,1)} - [J_\Omega(n_{E\bm{k}}^{(0)})]^{(1,1)}
\end{align}
here, we assume that the diagonal part of driving term is $D^{(d)}_{E \bm{k}} = D_0 \sigma_0 + D_{\Omega} \sigma_\Omega $. Thus, $\Delta_2=0$ yields us,
\begin{align}
     n_{E\bm{k}}^{(0)}  &= -\frac{\hbar e E_x }{ m} \tau k \cos(\theta)   \delta  \left(\epsilon _k^{(0)}-\epsilon _F\right)
     \nonumber \\
     n_{E\bm{k}}^{(\Omega)}  &= \frac{\tau e E_x}{\hbar} \cos(\theta) \bigg( \frac{\hbar^2 k}{m}  \bigg)  \left(\Omega_k \frac{\partial }{\partial \epsilon_{k}^{(0)}}\delta( \epsilon_{k}^{(0)}-\epsilon _F) \right)  + \frac{\tau e E_x}{\hbar} \cos (\theta ) \bigg( \frac{2 B^2 k^3  \Delta_1^2  }{\Omega_k} \bigg)  \delta  \left(\epsilon _k^{(0)}-\epsilon _F\right)  
     \label{diagonal-density}
\end{align}
and $\tau=\frac{\hbar^3}{n_i U^2 m}$ is the relaxation time. The disorder potential matrix elements in equation \ref{J_d_diagonal_density} are given by,

\begin{align}
    U_{\bm{k},\bm{k}'}{}^{(m,m')}= \langle m, \bm{k}|U| m', \bm{k}' \rangle
    \end{align}

\begin{align}
    & U_{\bm{k},\bm{k}'}{}^{(1,1)} = \frac{U}{2 \sqrt{\Omega_k \Omega_{k'}}}  \left(e^{+2 i (\theta - \theta')} \sqrt{\Omega_k -M} \sqrt{\Omega_{k'}-M} +  \sqrt{\Omega_k + M } \sqrt{\Omega_{k'}+M}\right)
    \\
    &  U_{\bm{k}',\bm{k}}{}^{(1,1)} = \langle 1,\bm{k}' |U| 1,\bm{k} \rangle = U \langle 1,\bm{k}' | 1,\bm{k} \rangle 
    = (U \langle 1,\bm{k} | 1,\bm{k}' \rangle)^{\dagger} = \left(U_{\bm{k},\bm{k}'}{}^{(1,1)}\right)^{\dagger}
\end{align}

\begin{align}
    & U_{\bm{k},\bm{k}' }{}^{(1,2)} = \frac{U}{2 \sqrt{\Omega_k  \Omega_{k'}}}
     \left(-e^{2 i (\theta - \theta')} \sqrt{\Omega_k -M} \sqrt{\Omega_{k'}+M}  +\sqrt{\Omega_k + M} \sqrt{\Omega_{k'}-M}\right)
     \\
    & U_{\bm{k}',\bm{k} }{}^{(2,1)} = \langle 2,\bm{k}' |U| 1,\bm{k} \rangle = U \langle 2,\bm{k}' | 1,\bm{k} \rangle  = [U \langle 1,\bm{k} | 2,\bm{k}' \rangle]^{\dagger} = \left( U_{\bm{k},\bm{k}' }{}^{(1,2)} \right)^{\dagger}     
\end{align}

\begin{align}
    \begin{split}
    & |U_{\bm{k},\bm{k}'}{}^{(1,1)}|^2   + |U_{\bm{k},\bm{k}'}{}^{(1,2)}|^2  = U^2
    \label{U(1,1)^2+U(1,2)^2}
\end{split}
\end{align}

\begin{align}
    \begin{split}
    & -|U_{\bm{k},\bm{k}'}{}^{(1,1)}|^2   + |U_{\bm{k},\bm{k}'}{}^{(1,2)}|^2   = \frac{U^2}{4 \Omega_k \Omega_{k'}}  \bigg\{- 4 M^2 - 2 \left(\sqrt{\Omega_k+M} \sqrt{\Omega_k -M} \sqrt{\Omega_{k'}+M} \sqrt{\Omega_{k'}-M} \right) 
    \\
    &   \left[e^{2 i (\theta - \theta')}  + e^{-2 i (\theta - \theta')} \right] \bigg\} 
    \label{-U(1,1)^2+U(1,2)^2}
\end{split}
\end{align}

\begin{align}
    & U_{\bm{k},\bm{k}'}{}^{(2,2)} =  \frac{U}{2 \sqrt{ \Omega_k \Omega_{k'}}}  \left( e^{2 i(\theta - \theta')}  \sqrt{\Omega_k + M} \sqrt{\Omega_{k'}+M} + \sqrt{\Omega_k-M}\sqrt{\Omega_{k'}-M} \right)
    \\
    & U_{\bm{k}',\bm{k}}{}^{(2,2)} =  \langle 2, \bm{k}'|U| 2, \bm{k} \rangle =  U\langle 2, \bm{k}' | 2, \bm{k} \rangle = (U\langle 2, \bm{k} | 2, \bm{k}' \rangle )^{\dagger} =(U_{\bm{k},\bm{k}'}{}^{(2,2)} )^{\dagger}
\end{align}

\begin{align}
    & U_{\bm{k},\bm{k}'}{}^{(2,1)} = \frac{U}{2 \sqrt{ \Omega_k \Omega_{k'}} }   \left( -e^{2 i \left(\theta- \theta'\right) } \sqrt{\Omega_k + M} \sqrt{\Omega_{k'}-M}  +  \sqrt{\Omega_k - M }  \sqrt{\Omega_{k'}+M}  \right) 
    \\
    &  U_{\bm{k}',\bm{k}}{}^{(1,2)}= \langle 1, \bm{k}'|U| 2, \bm{k} \rangle  =  U\langle 1, \bm{k}' | 2, \bm{k} \rangle  =\left( U \langle 2, \bm{k} | 1, \bm{k}' \rangle \right)^{\dagger} = \left( U_{\bm{k},\bm{k}'}{}^{(2,1)} \right)^{\dagger}
\end{align}

\begin{align}
\begin{split}
    & |U_{\bm{k},\bm{k}'}{}^{(2,1)}|^2 + |U_{\bm{k},\bm{k}'}{}^{(2,2)}|^2  = U^2
    \label{U(2,1)^2 + U(2,2)^2}
\end{split}
\end{align}

\begin{align}
    \begin{split}
        -|U_{\bm{k},\bm{k}'}{}^{(2,1)}|^2 + |U_{\bm{k},\bm{k}'}{}^{(2,2)}|^2  = \frac{U^2}{4 \Omega_{k} \Omega_{k'}}  \bigg\{ 4 M^2 + \left( 2\sqrt{\Omega_k +M} \sqrt{\Omega_k -M} \sqrt{\Omega_{k'}+M} \sqrt{\Omega_{k'}-M}\right)   \left[  e^{2 i (\theta-\theta') }   + e^{-2 i (\theta-\theta') }  \right]  \bigg\} 
    \end{split}
    \end{align}

Delta functions can be Taylor expanded in terms of pure kinetic parts and split energies,  
\begin{align}
    &\delta \left(\epsilon _{\bm{k}}^{(1)}-\epsilon _{\bm{k}'}^{(1)}\right) = \delta \left(\epsilon _{k}^{(0)}-\epsilon _{k'}^{(0)}\right)-(\Omega_{\bm{k}} -\Omega_{\bm{k'}}) \frac{\partial }{\partial \epsilon _{k}^{(0)}}\delta  \left(\epsilon _{k}^{(0)}-\epsilon _{k'}^{(0)}\right) \nonumber
    \\
    & \delta \left(\epsilon _{\bm{k}}^{(1)}-\epsilon _{\bm{k}'}^{(2)}\right) = \delta \left(\epsilon _{k}^{(0)}-\epsilon _{k'}^{(0)}\right)-(\Omega_{\bm{k}} +\Omega_{\bm{k'}}) \frac{\partial }{\partial \epsilon _{k}^{(0)}}\delta  \left(\epsilon _{k}^{(0)}-\epsilon _{k'}^{(0)}\right) \nonumber
    \\
    &\delta \left(\epsilon _{\bm{k}}^{(2)}-\epsilon _{\bm{k}'}^{(1)}\right) = \delta \left(\epsilon _{k}^{(0)}-\epsilon _{k'}^{(0)}\right) + (\Omega_{\bm{k}} +\Omega_{\bm{k'}}) \frac{\partial }{\partial \epsilon _{k}^{(0)}}\delta  \left(\epsilon _{k}^{(0)}-\epsilon _{k'}^{(0)}\right) \nonumber
    \\
    & \delta \left(\epsilon _{\bm{k}}^{(2)}-\epsilon _{\bm{k}'}^{(2)}\right) = \delta \left(\epsilon _{k}^{(0)}-\epsilon _{k'}^{(0)}\right) + (\Omega_{\bm{k}} -\Omega_{\bm{k'}}) \frac{\partial }{\partial \epsilon _{k}^{(0)}}\delta  \left(\epsilon _{k}^{(0)}-\epsilon _{k'}^{(0)}\right)
\end{align}

Across our calculations we have used widely the following properties. 
\begin{align}
    & \delta  \left(\epsilon_{k}^{(0)}-\epsilon _F\right) = \frac{\delta  \left( k - k_F\right)}{\frac{\hbar^2 k_F}{m}} \\
    & \frac{\partial }{\partial \epsilon_{k}^{(0)}} = \frac{1}{\frac{\hbar^2 k}{m}}\frac{\partial }{\partial k}
\end{align}

\begin{align}
    & k_x= k_x(k,\theta ), \quad k_y= k_y(k,\theta ) \\
    & \frac{\partial }{\partial k_x} = \cos (\theta ) \frac{\partial }{\partial k}-\frac{1}{k}\sin (\theta )\frac{\partial }{\partial \theta }
    \\
    & \frac{\partial }{\partial k_y} = \sin (\theta ) \frac{\partial }{\partial k} + \frac{1}{k}\cos (\theta )\frac{\partial }{\partial \theta }
\end{align}

\section{Appendix B: Second order response}

Driving term in the second order is given by,
\begin{align}
    & D_{E2} = \frac{e E_x}{\hbar}\left(\frac{\partial \rho_E }{\partial k_x}-i[\mathcal{R}_{k_x}, \rho_E ] \right) - \frac{i}{\hbar}\left[\tilde{H}_{\lambda }, \rho_E  \right]
\end{align}
in which $ \rho_E $ is the density matrix derived in linear response. Here we carry out very similar calculations as we carried on in the linear response to find out $\rho_{E2}$ the second-order density matrix response. Replacing $\rho_E$ by $\rho_0$ in equation \ref{differential_1st} and renaming $\rho_{E}$ to $\rho_{E2}$, we can work and find again the diagonal and off-diagonal parts of $\rho_{E2} = n_{E2} + S_{E2}$. The off-diagonal part of $\rho_{E2}$ can be calculated using equation \ref{S_E&D_E} in the following form,
\begin{align}
    S_{E2} & = S_{E2}^{(1)} \sigma_1 +S_{E2}^{(2)} \sigma_2
\end{align}
We find the $ S_{E2}$ in terms of $\sigma_1$ and $\sigma_2$,
\begin{align}
    & S_{E2} = \frac{e E_x}{4 \Omega ^6} \Big\{ -e E_x k \lambda  M \Omega_k ^2 \cos (\theta ) 
    \left[2 B^2 \Delta_1^2 k^4  - \Omega_k ^2 ( 2  +k \partial_k )\right](f_1-f_2)  \nonumber 
    \\
    &  + e E_x B \Delta_1 M \cos^2(\theta )  \left[6 B^2 \Delta_1^2 k^4 -\Omega_k^2 ( 1 + k \partial_k )\right] (f_1-f_2)  +B \Delta_1 \Omega_k^2 \sin (\theta ) \left[ e E_x M (f_1-f_2) \sin (\theta ) -4 k n_\Omega \Omega_k^2 \right] \Big\} \sigma_1 \nonumber
    \\
    &  + \frac{e E_x}{4 \Omega_k^5} \Big\{ 2 k \lambda  M \Omega_k^2 \left[e E_x M (f_1-f_2)  \sin (\theta )-2 k n_\Omega \Omega_k ^2\right]  \nonumber 
    \\
    & + B \Delta_1 \cos (\theta ) \left[ 4 k M n_\Omega \Omega_k^2 - e E_x \sin (\theta )    \left(-4 B^2 \Delta_1^2 k^4  +2 M^2 + k \Omega_k ^2 \partial_k  \right) (f_1-f_2) \right] \Big\} \sigma_2.
\end{align}
where $n_\Omega$ is given at equation \ref{diagonal-density}.
The diagonal parts of $\rho_{E2}$ becomes,
\begin{align}
    n_{E 2} = n_{E 2}^{(0)} \sigma_0 + n_{E2}^{(\Omega)} \sigma_3
\end{align}

\begin{align}
    n_{E 2}^{(0)} = & - \frac{ e^2 E_x^2 }{ m } \tau^2 \delta(\epsilon_k^{(0)} - \epsilon_F)  -\frac{ e^2 E_x^2 }{m} \tau^2 k \cos^2(\theta) \frac{\partial}{\partial k} \delta(\epsilon_k^{(0)} - \epsilon_F)
\end{align}

\begin{align}
    n_{E 2}^{(\Omega)} & =  \frac{2 e^2 E_x^2  (M^2 + \Omega_k^2) \tau^2}{k^2 \hbar^2 \Omega_k} \delta(\epsilon_k^{(0)} - \epsilon_F)  +  \frac{ e^2 E_x^2  \Omega_k (-7 M^2 + 8 \Omega_k^2 ) \tau^2}{2 k \hbar^2 (-M^2 + \Omega_k^2)} \frac{\partial}{\partial k} \delta(\epsilon_k^{(0)} - \epsilon_F) \nonumber \\
    & +  \frac{ e^2 E_x^2  \Omega_k (-2 M^2 + 4 \Omega_k^2 ) \tau^2}{4 \hbar^2 (-M^2 + \Omega_k^2)} \frac{\partial}{\partial k} \frac{\partial}{\partial k} \delta(\epsilon_k^{(0)} - \epsilon_F)  + \frac{ 4 e^2 E_x^2 M^2 (-M^2 + \Omega_k^2 ) \cos(2\theta) \tau^2}{k^2  \hbar^2 \Omega_k (M^2 + \Omega^2)}  \delta(\epsilon_k^{(0)} - \epsilon_F) \nonumber \\
    & +  \frac{ e^2 E_x^2  \Omega_k (-5 M^2 + 3 \Omega_k^2 ) \cos(2\theta) \tau^2}{2 k \hbar^2  (M^2 + \Omega_k^2)}  \frac{\partial}{\partial k}\delta(\epsilon_k^{(0)} - \epsilon_F)  +  \frac{ e^2 E_x^2  \Omega_k \cos(2\theta) \tau^2}{2 \hbar^2 }  \frac{\partial}{\partial k} \frac{\partial}{\partial k} \delta(\epsilon_k^{(0)} - \epsilon_F) 
\end{align}

Density matrix can be expressed in terms of diagonal and off-diagonal terms, $\rho_{E2} = n_{E2} + S_{E2}$.

\begin{align}
    & \text{tr}(v_x \rho_{E2}) = \text{tr} \left[v_x^{(d)} n_{E2} \right] + \text{tr} \left[ v_x^{(od)} S_{E2}  \right] \nonumber
    \\
    & \text{tr}(v_y \rho_{E2})  =   \text{tr} \left[v_y^{(d)} n_{E2} \right] + \text{tr} \left[ v_y^{(od)} S_{E2}  \right]
    \label{trace}
\end{align}
 where $(d)$ and $(od)$ denotes the diagonal and off-diagonal parts of the velocity matrices along x and y directions. Diagonal parts of velocities at $\Delta_2=0$ are given by,

\begin{align}
    &  v_{x}^{(d)} = \frac{1}{\hbar} \left(
    \begin{array}{cc}
    \frac{\partial \epsilon_k^{(1)}}{\partial k_x} & 0 \nonumber \\
    0 &  \frac{\partial \epsilon_k^{(2)}}{\partial k_x} \nonumber \\
    \end{array}
    \right)  = \frac{1}{\hbar}  \frac{\hbar^2 k}{m} \cos (\theta )  \sigma_0   - \frac{1}{\hbar} \frac{2 B^2 \Delta_1^2 k^3 \cos (\theta ) }{\Omega_{k} } \sigma_3
\end{align}

\begin{align}
    & v_{y}^{(d)} = \frac{1}{\hbar} \left(
    \begin{array}{cc}
    \frac{\partial \epsilon_k^{(1)}}{\partial k_y} & 0 \nonumber \\
    0 &  \frac{\partial \epsilon_k^{(2)}}{\partial k_y} \nonumber \\
    \end{array}
    \right) = \frac{1}{\hbar}  \frac{\hbar^2  k}{ m} \sin (\theta ) \sigma_0 - \frac{1}{\hbar} \frac{2 B^2 \Delta_1^2 k^3 \sin (\theta ) }{\Omega_{k} } \sigma_3
\end{align}
We express $H_0$ in its own eigenbasis,
\begin{align}
    & H_{0,diagonal}=\left(
    \begin{array}{cc}
        \epsilon_k^{(0)} - \Omega_{\bm{k}} & 0 \\ 
     0  & \epsilon_k^{(0)} + \Omega_{\bm{k}} \\
    \end{array}
    \right)  = \epsilon_k^{(0)} \sigma_0 - \Omega_{\bm{k}} \sigma_3 
\end{align}
The off-diagonal parts of velocity is given by, 
\begin{align}
    & v_{x}^{(od)} = -\frac{i}{\hbar}\left[\mathcal{R}_{k_x},H_{0,diagonal}\right]  =   \left( \frac{ 2 B k \sin (\theta ) \Delta_1}{\hbar   }\right)  (\sigma_2)  + \left( \frac{2 B k M \cos (\theta ) \Delta_1}{ \hbar  \Omega_{k}}\right)  (  \sigma_1)  
\end{align}

\begin{align}
    & v_{y}^{(od)} = -\frac{i}{\hbar}\left[\mathcal{R}_{k_y},H_{0,\text{diagonal}}\right]   =   \left( -\frac{ 2 B k \cos (\theta ) \Delta_1 }{\hbar   }  \right) ( \sigma_2)
    +  \left(\frac{2 B k M \sin (\theta ) \Delta_1}{\hbar \Omega_{k}}  \right) ( \sigma_1)
\end{align}
Now we can apply the traces at Eq. \ref{trace} by integrating over $(k,\theta)$ plane, 
 \begin{align}
    \int_{0}^{2\pi} \int_{0}^{\infty} \frac{(dk) (k d\theta) }{(2\pi)^2} v_{x}^{(d)}  n_{E2} = 0
 \end{align}

 \begin{align}
    \int_{0}^{2\pi} \int_{0}^{\infty} \frac{(dk) (k d\theta) }{(2\pi)^2} v_{y}^{(d)} n_{E2} = 0
 \end{align}

 \begin{align}
    \int_{0}^{2\pi} \int_{0}^{\infty} \frac{(dk) (k d\theta) }{(2\pi)^2} v_{x}^{(od)} S_{E2} = - \frac{2 e^2 E_x^2 B^3 \Delta_1^3 k_F^6 m M^2 \lambda}{\pi  \hbar^3 \left(B^2 \Delta_1^2 k_F^4+M^2 \right)^2} 
 \end{align}
 
 \begin{align}
    \int_{0}^{2\pi} \int_{0}^{\infty} \frac{(dk) (k d\theta) }{(2\pi)^2} v_{y}^{(od)} S_{E2} =   -\frac{2 e^2 E_x^2 \lambda m M \tau \left(B^3 \Delta_1^3 k_F^6 + 2 B \Delta_1 k_F^2 M^2\right)}{\pi h^4 \left(B^2 \Delta_1^2 k_F^4+M^2 \right)}
\end{align}

\section{Appendix C: Schrieffer-Wolff transformation}

We are going to determine the $\lambda$ value in equation \ref{H_lambda_1st}. To do so we need to solve Luttinger Hamiltonian along with $H_E$ (Eq. \ref{H_E_full}). we define the total angular momentum matrices for spin-3/2 in $\{3/2, -3/2, 1/2, -1/2\}$ basis, 
\begin{align}
J_x =    \left(
\begin{array}{cccc}
 0 & 0 & \frac{\sqrt{3}}{2} & 0 \\
 0 & 0 & 0 & \frac{\sqrt{3}}{2} \\
 \frac{\sqrt{3}}{2} & 0 & 0 & 1 \\
 0 & \frac{\sqrt{3}}{2} & 1 & 0 \\
\end{array}
\right)
\end{align}

\begin{align}
J_y = 
\left(
\begin{array}{cccc}
 0 & 0 & -\frac{1}{2} \left(i \sqrt{3}\right) & 0 \\
 0 & 0 & 0 & \frac{i \sqrt{3}}{2} \\
 \frac{i \sqrt{3}}{2} & 0 & 0 & -i \\
 0 & -\frac{1}{2} \left(i \sqrt{3}\right) & i & 0 \\
\end{array}
\right)
\end{align}

\begin{align}
J_z = 
\left(
\begin{array}{cccc}
 \frac{3}{2} & 0 & 0 & 0 \\
 0 & -\frac{3}{2} & 0 & 0 \\
 0 & 0 & \frac{1}{2} & 0 \\
 0 & 0 & 0 & -\frac{1}{2} \\
\end{array}
\right)
\end{align}

The terms corresponding to the applied electric field enter to the Hamiltonian in the following form,
\begin{align}
    & H_E = e a_B \xi E_x \{J_y,J_z\} + e a_B \xi E_y \{J_z,J_x\} 
    \label{H_E_full}
\end{align}
in which the $a_B$ is the Bohr radius and $\xi$ is a parameter controlling the intensity of electric-dipole terms. The Zeeman Hamiltonian is given as,

\begin{align}
    &H_B = \bm{B}.\bm{J} = B_x J_x+ B_y J_y + B_z J_z
\end{align}

The Luttinger Hamiltonian in the same basis that the total angular momentum matrices are provided, becomes,

\begin{align}
    & H_L=  \left(
\begin{array}{cccc}
 \varepsilon_h+\frac{3}{2}M & 0 & L & R \\
 0 & \varepsilon_h-\frac{3}{2}M & R^{*} & -L^{*} \\
 L^{*} & R & \varepsilon_l+\frac{1}{2}M & 0 \\
 R^{*} & -L & 0 & \varepsilon_l-\frac{1}{2}M \\
\end{array}
\right)
\end{align}
\begin{align}
    & \varepsilon_h = P+Q && \varepsilon_l = P-Q \nonumber \\
    &  \bar{\gamma} =\frac{ \gamma_2 + \gamma_3}{2}  && \zeta = \frac{ \gamma_3 - \gamma_2}{2}  \nonumber \\
    & k_\pm = k_x \pm i k_y &&    P = \frac{1}{2} \gamma_1 \mu  \left(k^2+ k_z^2\right) \nonumber \\
    & k^2=k_x^2+ k_y^2 &&  L = -\sqrt{3} \mu \gamma_3 k_- k_z   &&  \nonumber \\
    &  Q=-\frac{1}{2} \gamma_2 \mu \left(2 k_z^2-k^2 \right) && R=-\frac{1}{2} \left(\sqrt{3} \mu \right) \left(\bar{\gamma} k_-^2 - \zeta  k_+^2 \right)
\end{align}

We use Schrieffer-Wolff transformation to project $H_L + H_E$ to the HH subspace. To carry on the transformation we need to write $H_L + H_E = H^0 + H'$, where $H^0$ contains the diagonal terms and $H'$ includes the the block off-diagonal terms. The transformed Hamiltonian in HH basis is expressed as,  
\begin{align}
    \tilde{H} = H^{(0)} + H^{(1)} + H^{(2)} + \cdots
\end{align}
where 
\begin{align}
    & H^{(0)}_{m m'} = H^{0}_{m m'} \\
    & H^{(1)}_{m m'} = H'_{m m'} \\
    & H^{(2)}_{m m'} = \frac{1}{2} \sum_{l} H'_{ml} H'_{lm'} \left(\frac{1}{E_m - E_l} + \frac{1}{E_{m'} - E_l} \right)  
\end{align}
where $E_m$, $E_{m'}$, and $E_l$ takes values from $H^0$ for $m,m'=1,2$ and $l=3,4$. The off-diagonal terms of $\tilde{H}$ becomes,
\begin{align}
    \tilde{H}_{od} = \frac{3 a_B \gamma_3 e E_x k_x k_y M \xi }{4 \gamma_2^2 \langle k_z^2 \rangle ^2 \mu } \sigma_1 + \frac{3 a_B e E_x M \xi  \left(k_y^2-k_x^2\right)}{8 \gamma_2 \langle k_z^2 \rangle ^2 \mu } \sigma_2 
\end{align}
Comparing the off-diagonal terms of $\tilde{H}$ with $ H_\lambda = 2 e E_x \lambda  k_x k_y M \sigma_1 - e E_x \lambda  M \left(k_x^2 - k_y^2 \right) \sigma_2$,

\begin{align}
    & \frac{3 a_B \gamma_3 e E_x k_x k_y M \xi }{4 \gamma_2^2 \langle k_z^2 \rangle ^2 \mu } \sigma_1 = 2 e E_x \lambda  k_x k_y M \sigma_1 \\
    & \text{and} \nonumber \\
    & \frac{3 a_B e E_x  \left(k_y^2-k_x^2\right) M \xi}{8 \gamma_2 \langle k_z^2 \rangle ^2 \mu } \sigma_2 = - e E_x \lambda  M \left(k_x^2 - k_y^2 \right) \sigma_2
\end{align}
in axial approximation both equations are reduced to,
\begin{align}
    \lambda = \frac{3 a_B \xi}{8 \bar{\gamma} \langle k_z^2 \rangle^2 \mu}
\end{align}

Diagonal part of the transformed Hamiltonian are given as,

\begin{align}
    H_{d}= H_{d}^{(0)} \sigma_0 +H_{d}^{(3)} \sigma_3 
\end{align}

\begin{align}
    & H_{d}^{(0)}=  -\frac{3 \left(B_x^2+B_y^2\right)}{8 \gamma_2 k_z^2 \mu } - \frac{3 a_B e \xi  (B_z+M) (B_x E_y+B_y E_x)}{4 \gamma_2^2 k_z^4 \mu ^2} -\frac{3 \gamma_2 \mu  \left(k_x^2+k_y^2\right)^2}{8 k_z^2} -\frac{3 a_B^2 e^2 \xi ^2 \left(E_x^2+E_y^2\right)}{2 \gamma_2 k_z^2 \mu }
     +\frac{1}{2} k_z^2 \mu  (\gamma_1-2 \gamma_2)
\end{align}

\begin{align}
    & H_{d}^{(3)}=  -\frac{3 (B_z+M) \left(k_x^2+k_y^2\right)^2}{8 k_z^4} -\frac{3 a_B e \xi  (B_x E_y+B_y E_x)}{2 \gamma_2 k_z^2 \mu } -\frac{3 (B_z+M) \left(4 a_B^2 e^2 \xi ^2 \left(E_x^2+E_y^2\right)+B_x^2+B_y^2\right)}{16 \gamma_2^2 k_z^4 \mu ^2}  +\frac{3 (B_z+M)}{2}
\end{align}

\section{Appendix D: Nonlinear Hall voltage}

We are interested to see what is the order of Hall potential in our system for a sample size of $9.2 \, \mu m$ in $\hat{\bm{y}}$ direction \cite{kang2019nonlinear},

\begin{align}
    \left(
\begin{array}{cc}
 \rho _{xx} & \rho _{xy} \\
 \rho _{yx} & \rho _{yy} \\
\end{array}
\right)=\frac{1}{\det (\sigma) } \left(
    \begin{array}{cc}
     \sigma _{yy} & -\sigma _{xy} \\
     -\sigma _{yx} & \sigma _{xx} \\
    \end{array}
    \right)=\frac{1}{\sigma _{xx}^2+\sigma _{yx}^2} \left(
        \begin{array}{cc}
         \sigma _{xx} & -\sigma _{xy} \\
         \sigma _{xy} & \sigma _{xx} \\
        \end{array}
        \right)
\end{align}

Resistivity and conductivities corresponding to linear responses are given by $\rho$ and $\sigma$ matrices. 

\begin{align}
    E_y = j_x \rho _{yx}+j_y \rho_{yy}= \frac{\sigma _{xy}}{\sigma _{xx}^2+\sigma _{yx}^2} ( \sigma _{xx} E_x + \chi _{xxx} E_x^2) + \frac{\sigma _{xx}}{\sigma _{xx}^2+\sigma _{yx}^2} (\sigma _{yx} E_x + \chi _{yxx} E_x^2) 
\end{align}
We have expanded the current densities to include linear and second-order responses. Considering that $\sigma_{xx} \gg \sigma_{xy}$, the induced electric field across y-axis in second-order regime becomes,
\begin{align}
    E_y^{(2)} \approx \frac{ \chi _{yxx}}{\sigma _{xx}} E_x^2
\end{align}

%I applied the blue colored M=0.1meV for your attention
$M=0.1meV$:
From figure \ref{fig:3D} we take the current density $0.96 \, \mu A/m$ corresponding to the magnetic field, electric field, and magnetization provided at $B= 2 \, T$, $E=30 \, kV/m$, and $M=0.1 \, meV$. To find the conductivity we consider that the relaxation time is in order of picosecond. The hole carriers density can be taken as $1.86 \times 10^{15} \, m^{-2}$. The HH mass is $m^* = m_0 / (\gamma_1+ \bar{\gamma})$.
\begin{align}
    \sigma _{xx} = \frac{n e^2 \tau}{ m} = \frac{(1.86 \times 10^{15}) (1.602 \times 10^{-19})^2 (1\times 10^{-12})}{9.109 \times 10^{-31} / 9.35 } = 4.91 \times 10^{-4}  \frac{1}{\Omega}\approx 12.7 \, \frac{e^2}{h}.
\end{align}
where $\Omega$ is simply ohm. Therefore the induced electric field is
\begin{align}
    E_y^{(2)} \approx \frac{ \chi _{yxx}}{\sigma _{xx}} E_x^2 = \frac{9.66 \times 10^{-7} }{4.91 \times 10^{-4} } = 1.96 \times 10^{-3} \, V/m
\end{align}
yielding the non-linear Hall voltage
\begin{align}
    V_y^{(2)} = E_y^{(2)} d = 1.96 \times 10^{-3} \times 9.2 \times 10^{-6}= 18.032 \times 10^{-9} \, V = 18.032 \, {\rm n V}. 
\end{align}

$M=2$ meV: From figure \ref{fig:3D_2nd} we take the current density $0.249 \, m A/m$ corresponding to the magnetic field, electric field, and magnetization provided at $B= 2 \, T$, $E=30 \, kV/m$, and $M=2 \, meV$. To find the conductivity we consider that the relaxation time is in order of picosecond. The hole carriers density can be taken as $1.86 \times 10^{15} \, m^{-2}$. The HH mass is $m^* = m_0 / (\gamma_1+ \bar{\gamma})$. The longitudinal conductivity is as before
\begin{align}
    \sigma _{xx} = \frac{n e^2 \tau}{ m} = \frac{(1.86 \times 10^{15}) (1.602 \times 10^{-19})^2 (1\times 10^{-12})}{9.109 \times 10^{-31} / 9.35 } = 4.91 \times 10^{-4} \frac{1}{\Omega} \approx 12.7 \, \frac{e^2}{h}.
\end{align}
Hence the induced electric field is
\begin{align}
    E_y^{(2)} \approx \frac{ \chi _{yxx}}{\sigma _{xx}} E_x^2 = \frac{2.49 \times 10^{-4} }{4.91 \times 10^{-4} } = 0.5 \, V/m
\end{align}
and the non-linear Hall voltage
\begin{align}
    V_y^{(2)} = E_y^{(2)} d = 0.50 \times 9.2 \times 10^{-6}= 4.6 \times 10^{-6} \, V = 4.6 \, \mu {\rm V}. 
\end{align}
\end{widetext}

\end{document}